\documentclass[conference]{IEEEtran}
\IEEEoverridecommandlockouts
\usepackage{cite}
\usepackage{amsmath,amssymb,amsfonts}
\usepackage{algorithmic}
\usepackage{graphicx}
\usepackage{textcomp}
\usepackage{xcolor}
\usepackage{colortbl}
\usepackage{multirow}
\usepackage{booktabs}
\usepackage{makecell}
\usepackage{subfigure}
\usepackage{multirow}

\def\BibTeX{{\rm B\kern-.05em{\sc i\kern-.025em b}\kern-.08em
    T\kern-.1667em\lower.7ex\hbox{E}\kern-.125emX}}
\begin{document}

\title{A Lightweight Approach for Network Intrusion Detection based on Self-Knowledge Distillation
\thanks{\IEEEauthorrefmark{2}Contributed Equally}
\thanks{\IEEEauthorrefmark{3}Corresponding Author}
}

\author{
\IEEEauthorblockN{
Shuo Yang\IEEEauthorrefmark{1}\IEEEauthorrefmark{2},
Xinran Zheng\IEEEauthorrefmark{1}\IEEEauthorrefmark{2}, Zhengzhuo Xu\IEEEauthorrefmark{1}, and
Xingjun Wang\IEEEauthorrefmark{1}\IEEEauthorrefmark{3}}
\IEEEauthorblockA{\IEEEauthorrefmark{1}Tsinghua Shenzhen International Graduate School, Tsinghua University, Shenzhen, China
}}
\maketitle

\begin{abstract}
Network Intrusion Detection (NID) works as a kernel technology for the security network environment, obtaining extensive research and application. Despite enormous efforts by researchers, NID still faces challenges in deploying on resource-constrained devices. To improve detection accuracy while reducing computational costs and model storage simultaneously, we propose a lightweight intrusion detection approach based on self-knowledge distillation, namely LNet-SKD, which achieves the trade-off between accuracy and efficiency. Specifically, we carefully design the DeepMax block to extract compact representation efficiently and construct the LNet by stacking DeepMax blocks. Furthermore, considering compensating for performance degradation caused by the lightweight network, we adopt batch-wise self-knowledge distillation to provide the regularization of training consistency. Experiments on benchmark datasets demonstrate the effectiveness of our proposed LNet-SKD, which outperforms existing state-of-the-art techniques with fewer parameters and lower computation loads.
\end{abstract}


\section{Introduction}

Accompanied by the rapid development of network technology, various network attacks emerge with more serious and huge threats \cite{networksecurity}. As a response, Network Intrusion Detection (NID) plays an essential role in providing the desired security by constantly monitoring malicious and suspicious activities in network traffic. Nowadays, Intrusion Detection Systems (IDSs) have been widely used in military, medical, transportation, IoT, industrial control systems, and other fields\cite{surveyappication}.

IDSs apply two types of detection manners, \textit{signature-based} and \textit{anomaly-based} \cite{surveytype}. The signature-based NID establishes the knowledge base by state modeling or string matching in advance and detects abnormal behavior by matching the data flow with the existing signature. Signature-based NID shows quite well performance on known attacks while failing to deal with attacks that are not in the knowledge base. Compared with it, anomaly-based NID has the ability to recognize unknown attacks by measuring the deviation between the detected activity and normal ones, which is vigorously developing.

\begin{figure}[t!]
\centering
\includegraphics[width=0.95\linewidth]{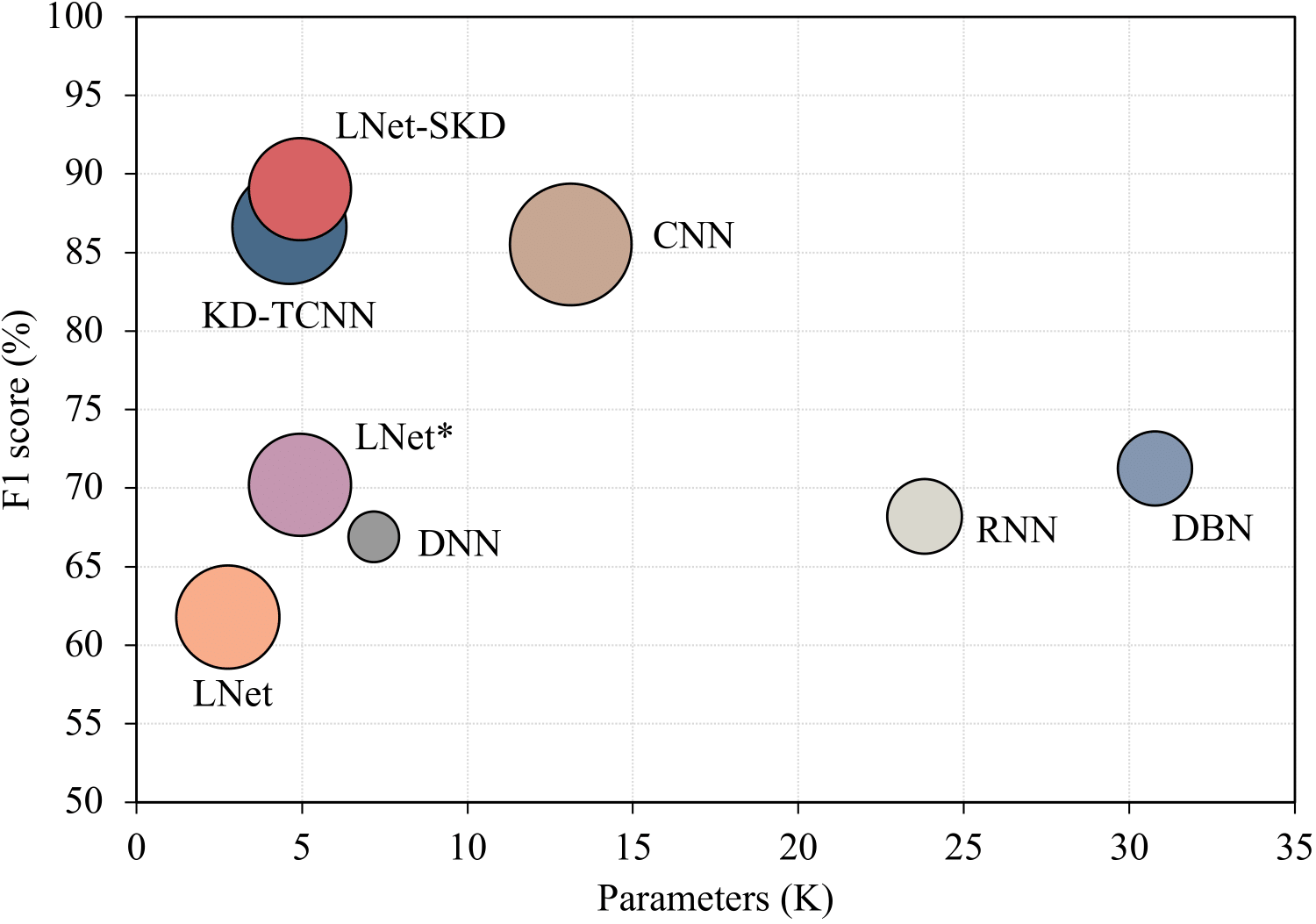}
\caption{F1 score \textit{v.s.} the number of model parameters on NSL-KDD datasets. Each data point is visualized as a circle whose radius is proportional to $log(p)$, where $p$ is the model's FLOPs. Notice that LNet-SKD achieves the best performance with satisfying model parameters and FLOPs.}
\label{teaser}
\end{figure}

With the success of Deep Learning (DL), numerous DL-based intrusion detection models \cite{CNN, bilstm, DBN} have been proposed and promote the accuracy and robustness of intrusion detection by a large margin. Despite the satisfactory accuracy these models achieved, most of them are difficult to be implemented on resource-constraint devices as high computational overhead and large model size. For example, DBN-based methods and RNN-based methods need more parameters resulting in the burden of storage (See Fig.~\ref{teaser}).

Several schemes have been proposed to lighten the model size for NID \cite{zhaoIOTJ, light2, light1}, where \cite{zhaoIOTJ} designed a lightweight network using depthwise convolution instead of standard convolution. However, the reduction of network complexity is at the expense of ignoring the correlation of channels, which leads to a decline in detection accuracy. In contrast, Depthwise Separable Convolution (DSConv) \cite{mobilenets} is more wildly used that introduces point-wise convolution to combine features between channels. Furthermore, motivated by the observation of the feature combination and selection ability of Max-Feature-Map (MFM) \cite{MFM}, we present a simple yet novel network structure named DeepMax block composed of DSConv and MFM, which allows a NID model to reinforce representation learning with lower computational cost.

To compensate for the performance degradation caused by the lightweight, an intuitive solution uses Knowledge Distillation (KD) \cite{KD, zhaoKD, KD-TCNN} to optimize the shallow student models by learning knowledge extracted from large and deep teacher models. However, we notice that student performance is highly dependent on the teacher models, which requires extra network design and results in additional training burdens to NID. To ameliorate it, we apply Self-Knowledge Distillation (SKD) \cite{skd-batch} to obtain the instantaneous knowledge generated during the training phase and improve the performance of lightweight models succinctly and effectively without heavy networks.

Based on the above analysis, we propose LNet-SKD for NID, which is a lightweight approach composed of LNet and SKD. LNet is a succinct but effective model stacked by several lightweight DeepMax blocks for the feature process. Furthermore, batch-wise SKD is employed to provide the regularization of training consistency, which improves the detection capability significantly. Compared with existing methods, LNet-SKD achieves superior performance with the trade-off between efficiency and accuracy (See Fig. \ref{teaser}). Our contributions are summarized as follows:

\begin{enumerate}
\item[1)] We propose the DeepMax with a meticulous design which is composed of MFM and DSConv, to reduce the model complexity while achieving efficient feature extraction and compact representation.
\item[2)] We propose the LNet by stacking DeepMax blocks for NID, which realizes satisfactory performance with lower model storage and computational cost. 
\item[3)] We utilize SKD to guide the LNet to obtain more instantaneous and coherent knowledge. To the best of our knowledge, we are the first to use SKD in NID to cover the performance drop incurred by lightweight.
\item[4)] Extensive experiments have shown that our LNet-SKD has significant advantages in accuracy and efficiency on challenging NSL-KDD and CICIDS-2017 datasets compared to state-of-the-art methods.
\end{enumerate}

\section{Related work}

Network intrusion detection is one of the most effective approaches for network security defense. Traditional IDSs are based on fixed or dynamic rules to detect network attacks, which are only suitable for relatively simple scenarios and are difficult to deal with unknown security risks \cite{signature}. With the development of Machine Learning (ML), almost all related algorithms have been applied to NID, such as KNN \cite{KNN}, SVM \cite{SVMBayes}, and LightGBM \cite{LightGBM}. However, they can no longer resist the increasingly complex and diverse network threats due to limited learning ability \cite{shallow}.

In recent years, DL has shown promising potential in learning inherent rules and the representation of samples and has been used to extract features from abnormal traffic in an end-to-end manner. \cite{CNN} proposed a deep learning approach for intrusion detection using the Convolution Neural Network (CNN). The proposed method effectively recognized abnormal traffic with higher accuracy than ML-based IDSs. However, it performed poorly against the minority classes in multi-class classification. Imrana \textit{et al.} \cite{bilstm} proposed a bidirectional Long-Short-Term-Memory (BiDLSTM) based intrusion detection system to improve the detection rate of minority classes. D'Angelo \textit{et al.} \cite{AutoRNN} embedded the autoencoder into the convolution neural network and recurrent neural network to obtain elicit relevant knowledge about the relations existing among the spatial features and temporal features, which are used to help improve the performance for network traffic classification. Belarbi \textit{et al.} \cite{DBN} developed a multi-class classification IDS based on Deep Belief Network (DBN) by stacking multiple Restricted Boltzmann Machines (RBMs), and its performance has been verified with the CICIDS2017 dataset. 

Complex networks provide high detection accuracy for IDS but bring challenges to deployment with resource-constrained devices. To address this problem, \cite{zhaoIOTJ} designed a lighter model by modifying the existing paradigm, which results in a loss of detection rate. Another popular solution is knowledge distillation. Wang \textit{et al.} \cite{KD-TCNN} proposed a knowledge distillation model to reduce the complexity of the model. Although the teacher model does improve the performance of the student model, it is challenging to design appropriate teacher and student models. Instead, our approach does not require a teacher model, distillation is conducted batch-wise, in which soft knowledge is propagated batch by batch.

\section{The proposed approach}

\begin{figure*}
\centering
\includegraphics[width=18cm]{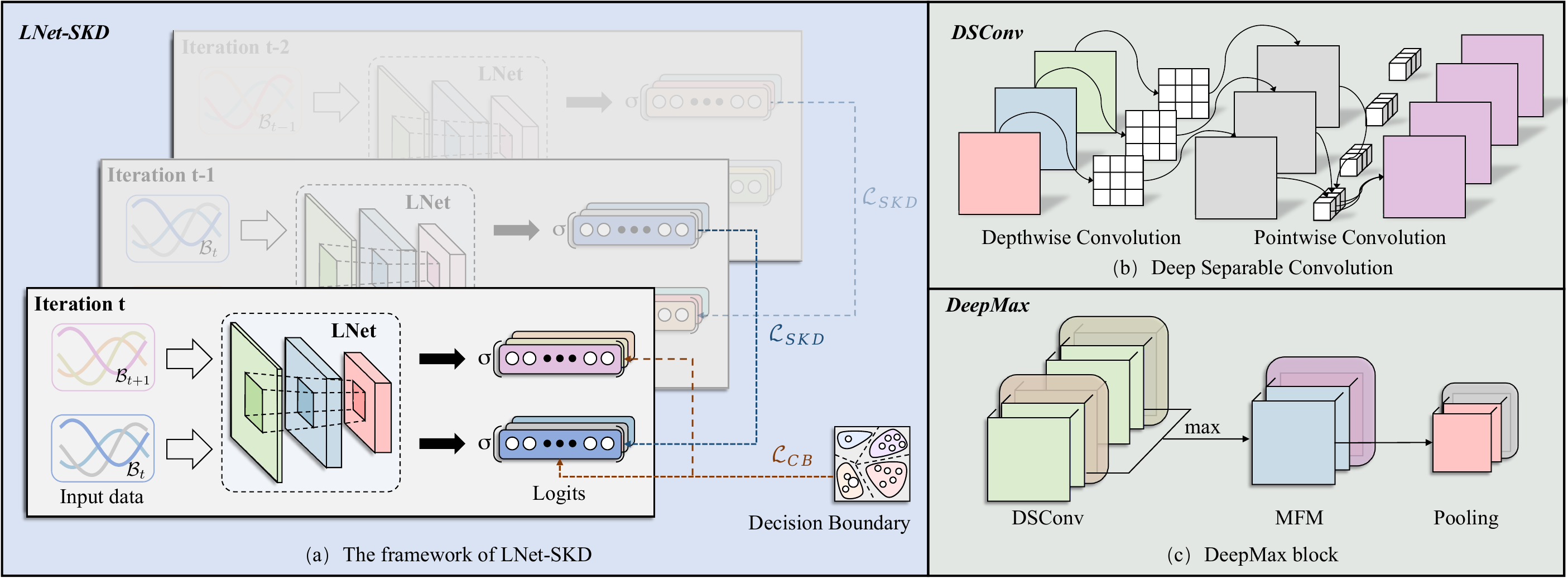}
\caption{The overview of proposed LNet-SKD for network intrusion detection. (a) The soft prediction in the last iteration will guide current iteration training. (b) Each filter kernel only calculates with the corresponding feature map. (c) Channels and feature sizes will be half with the DeepMax block.}
\label{approach}
\end{figure*}

In this section, we present a detailed discussion of the proposed lightweight self-knowledge distillation approach for network intrusion detection. Fig.~\ref{approach}(a) displays the framework of our approach, which comprises two essential parts. First, we propose the LNet by stacking lightweight DeepMax blocks which are carefully designed for feature extraction and selection without redundant parameters. LNet is able to extract the robust feature representation of intrusion behavior with lower computation overhead. Second, batch-wise self-knowledge distillation is introduced to instruct the LNet to acquire instantaneous and effective knowledge.

\subsection{Preliminary}
Given an $M$-classes labeled dataset containing $N$ training instances, $\mathcal{D}=\left\{\left(x_1, y_1\right),\left(x_2, y_2\right), \ldots,\left(x_N, y_N\right)\right\}$. Let $
\mathcal{B}_{t}=\left\{\left(x_1^{t}, y_1^{t}\right),\left(x_2^{t}, y_2^{t}\right), \ldots,\left(x_n^{t}, y_n^{t}\right)\right\}$ be a batch set of $t^{th}$ iteration during the training process, where $n \ll N$. Furthermore, we define a base NID model as ${\mathcal{M}_\theta}$, which is parameterized by $\theta$. For each input sample $(x,y)$, the encoder extracts the feature representation $\mathcal{F} \in \mathbb{R}^{H \times W}$, where $W$ and $H$ denote the spatial width and height of the corresponding feature map. We note the output logits ${\boldsymbol{z}=\mathcal{M}(x|\theta) \in \mathbb{R}^{M}}$. In this paper, both teacher and student adopt the same architecture ${\mathcal{M}_\theta}$ for self-knowledge distillation.

\subsection{LNet}
In NID, CNNs have been widely used due to their excellent feature extraction ability. Our motivation comes from the observation that the CNN-based models proposed for NID have constantly been growing larger, which hinders the deployment of edge devices. Hence, we propose an efficient feature processing block DeepMax and further build the lightweight LNet model by stacking it, as shown in Fig.~\ref{approach}. Under this structure, the computing overhead can be significantly reduced, and extract features effectively with generalization.

The core component of LNet is the DeepMax block, which is based on  \textit{Depthwise Separable Convolution (DSConv) layer} and \textit{Max-Feature-Map (MFM) layer}. As illustrated in Fig.~\ref{approach}(b), DSConv factorizes the standard convolution into depth-wise convolution and point-wise convolution. The depth-wise convolution operation extracts each channel's feature with separate kernels, which reduces the computation drastically. Then, point-wise convolution is used to match the output feature channel, which is implemented as $1\times1$ kernel convolution. The whole process for $C_i$ input and $C_o$ output channels can be written as:
\begin{equation}
    \mathcal{F}_{j}^{new} = \frac{1}{C_i} \cdot \mathcal{K}^{p}_{j} * \sum_{i=0}^{C_i-1} \mathcal{K}^{d}_{i} * \mathcal{F}_{i}^{old}, j \in [0,...,C_o-1],
\end{equation}
where $*$ indicates convolutional operation, $\mathcal{K}$ is the corresponding kernel and $\mathcal{F}$ is the feature map.

To further reduce the network size, we concatenate the MFM layer after the aforementioned DSConv layer to half the channel numbers. As shown in Fig.~\ref{approach}(c), MFM applies Eq.~\ref{MFM} to combine two feature maps and output the maximum value of each element, thus focusing the network on prominent elements.

\begin{equation}
\mathcal{F}_{i}^{new}(x,y) = \max \left[\mathcal{F}_{i}^{old}(x,y),  \mathcal{F}_{i+C/2}^{old}(x,y)\right],
\label{MFM}
\end{equation}
where $i \in [0, C/2-1]$ and $C$ is current number of channels.

Finally, we add a pooling layer to reduce the feature map size and construct the complete DeepMax block. Through effective feature extraction and selection, compact representation can be obtained with satisfactory computational consumption. To be implemented as a lightweight network for intrusion detection, it is enough to stack only two DeepMax blocks with a liner layer in LNet to achieve state-of-the-art results.

\subsection{Complexity analysis}
Here, we give an in-depth analysis of how LNet reduces the model size and saves the FLOPs. Consider an input feature map $\mathcal{F} \in \mathbb{R}^{H \times W}$ and in / out channel $C_{i}$ / $C_{o}$. In the standard convolution block, the filtering and combination steps are performed upon all input channels, and the number of parameters is $C_{i}\times K\times K\times C_{o}$, where $K$ denotes the kernel size. In contrast, our DeepMax block conducts separable convolution for each input channel and an additional $1\times1$ convolution to create a linear combination to match the output channel. Considering no extra parameters are introduced in the MFM or pooling operation, the total parameter number in the whole block is $C_{i}\times K\times K+C_{i}\times 1\times 1\times C_{o}$. It is clear that the DeepMax only needs $C_{i}\times (K^2+C_{o})$ gradient-required parameters, which is less than classical ones $C_{i}\times (K^2\times C_{o})$ by a large margin. For the computational cost, the DeepMax block narrows half the channel number by the MFM module and half the output feature size by the pooling layer if the pooling size is $2\times2$.

\subsection{Self-Knowledge Distillation}

To compensate for the performance drop incurred by model compression, we adopt the batch-wise self-knowledge distillation strategy inspired by \cite{skd-batch}, which improves the generalization ability by learning sample-level soft labels given by skillful teacher models. Specifically, the soft prediction of the last iteration is used as the smooth label for self-distillation to provide instantaneous distillation for each batch of training samples, which leads to the regularization of training consistency. As shown in Fig.~\ref{approach}(a), for each input sample$(x,y)$, the LNet will produce the predicted distribution $\boldsymbol{p}=\left\{p_1, \cdots, p_m \right\} \in \mathbb{R}^M$ by the suggested softmax function\cite{KD}.

\begin{equation}
p_i(x;\tau)=\frac{\exp \left(z_i(\mathbf{x}) / \tau\right)}{\sum_j \exp \left(z_j(\mathbf{x}) / \tau\right)},
\end{equation}
where $\tau$ denotes the temperature scale to soften the probability distribution for better distillation. Considering the imbalance of intrusion detection datasets, we use a class-balanced cross-entropy loss function to improve the attention to tail-category samples, which is written as follows:
\begin{equation}
\mathcal{L}_{\text {CB}}(\boldsymbol{z}, y)=-\frac{1-\beta}{1-\beta^{n_{y}}} \log \left(\frac{\exp \left(z_{y}/\tau\right)}{\sum_{j=1}^{M} \exp \left(z_{j}/\tau\right)}\right).
\end{equation}

Batch-wise distillation transfers the knowledge by optimizing the Kullback-Leibler (KL) divergence between the two batches' consecutive iterations and the loss of self-knowledge distillation as follows:

\begin{equation}
    \begin{aligned}
      \mathcal{L}_{SK D}&=\frac{1}{n} \sum_{i=1}^n \tau^2 \cdot D_{K L}\left(\boldsymbol{p}_i^{\tau, t-1} \| \boldsymbol{p}_i^{\tau, t}\right) \\
        &=\frac{1}{n} \sum_{i=1}^n \tau^2 \boldsymbol{p}_i^{\tau, t-1} \log(\boldsymbol{p}_i^{\tau, t-1}) - \boldsymbol{p}_i^{\tau, t-1}\log(\boldsymbol{p}_i^{\tau, t}),
    \end{aligned}
\end{equation}
where $\boldsymbol{p}_i^{\tau, t-1}$ is the soften labels generated by LNet at $(t-1)^{th}$ iteration and $\boldsymbol{p}_i^{\tau}$ is the soften predictions at $t^{th}$ iteration. The soften degree is controlled by temperature parameter $\tau$, higher temperatures lead to a more uniform distribution, resulting in a smoother batch-wise regularization effect. Compared to vanilla KD \cite{KD}, SKD plays a double role of student and teacher in the training phase. It keeps soft targets and extracts such smooth labels from the previous iteration for regularization. We use the soft prediction of the previous iteration to generate a dynamic sample-level smoothing label for self-distillation to provide the most instantaneous distillation for each training sample.

Combined the class-balanced loss and the SKD loss with a trade-off factor $\lambda$, we derive the overall loss function:
\begin{equation}
\mathcal{L}=\mathcal{L}_{CB}+\lambda \cdot \mathcal{L}_{SKD}.
\label{loss}
\end{equation}

\section{Experimental Results}

\begin{table}[!t]
\caption{NSL-KDD Dataset Description.}
\centering
\begin{tabular}{l|ccccc|c}
\toprule
Category & Normal & Dos   & Probe & R2L  & U2R & Total  \\ \midrule
Samples  & 77054  & 53385 & 14077 & 3749 & 252 & 148517 \\ \bottomrule
\end{tabular}
\label{Tab.NSL-KDD}
\end{table}
\begin{table}[!t]
\caption{CICIDS2017 Dataset Description.}
\centering
\resizebox{\linewidth}{5.5mm}{
\begin{tabular}{c|cccccc|c}
\toprule
Category & Benign  & DoS/DDoS & P.S. & B.F. & W.A. & Botnet & Total  \\ \midrule
Samples  & 2035505 & 320469   & 57305    & 8551        & 2118       & 1943   & 2425891 \\ \bottomrule
\end{tabular}
}
\label{Tab.CICIDS}
\end{table}

\begin{figure}
\subfigure[The impact of Temperature.]{
\includegraphics[width=0.9\linewidth]{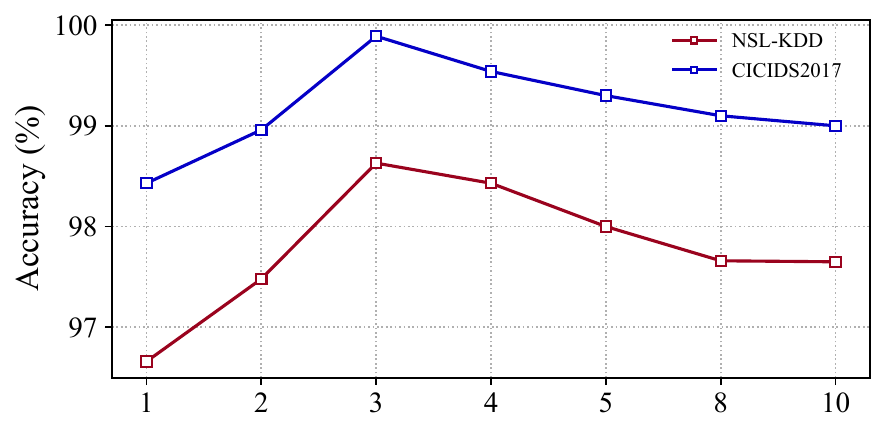} 
\centering
\label{T}
}
\subfigure[The impact of Lambda.]{
\includegraphics[width=0.9\linewidth]{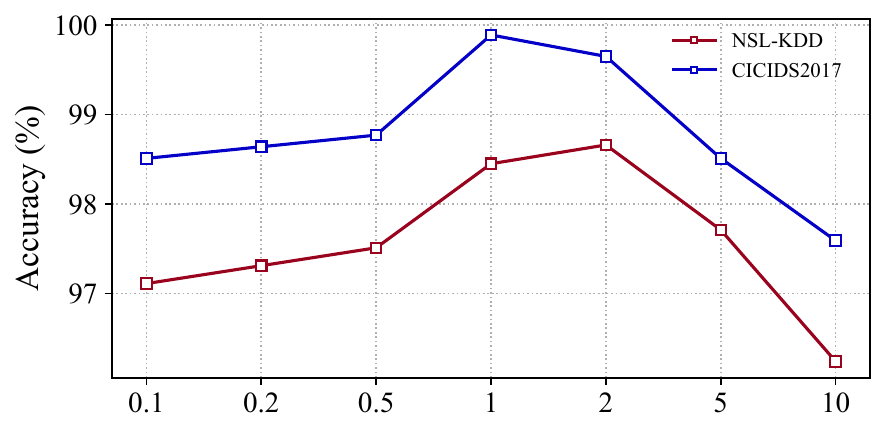} 
\centering
\label{L}
}
\DeclareGraphicsExtensions.
\caption{The impact of hyper-parameters.}
\label{hp}
\end{figure}

In this section, we first introduce the datasets and evaluation metrics used in experiments. All experiments are based on python 3.7 and PyTorch 1.12.0, using a 2.4GHz Intel Core i9 processor and 16GB RAM. We adopt stochastic gradient descent as the optimizer with a momentum of 0.9, and weight decay of $1e-4$. The initial learning rate is 0.1 and adjusted with cosine annealing.

\subsection{Dataset}
We use two benchmark datasets in experiments. One is the classic NSL-KDD dataset \cite{NSL-KDD} and the other is CICIDS2017 dataset \cite{CICIDS} which includes up-to-date typical attacks in the real world.

NSL-KDD dataset consists of 39 different types of attacks, which are divided into four main classes, i.e., Denial of Service (DoS), User-to-Root (U2R), Remote-to-Local (R2L), and Probe. Each intrusion record is composed of 9-\textit{dim} basic TCP connection features, 13-\textit{dim} TCP connection content features, 9-\textit{dim} time-based network traffic statistics features, 10-\textit{dim} host-based network traffic statistics features and the category label. The NSL-KDD dataset is described in Table \ref{Tab.NSL-KDD}.

CICIDS2017 dataset contains five days of network traffic data collected from Monday to Friday, including normal traffic and abnormal traffic caused by common attacks. The benign traffic corresponds to the human interaction of 25 users based on standard network protocols such as HTTP(S), FTP, SSH, IMAP, and POP3. Each record contains 6 basic features and more than 70 functional features. We follow previous work \cite{DBN} to utilize the dataset, and a detailed description of the CICIDS2017 dataset is shown in Table \ref{Tab.CICIDS}.

\begin{table*}[!ht]
\caption{Ablation with regard to LNet-SKD.}
\centering
\begin{tabular}{c|l|cc|cc|cc|cc|cc}
\toprule
Dataset                     & Model    & Acc.(\%) & $\Delta$(\%) & Prec.(\%) & $\Delta$(\%) & Recall(\%) & $\Delta$(\%) & F1(\%) & $\Delta$(\%) & Para.(K) & FLOPs(K) \\ \midrule
\multirow{4}{*}{NSL-KDD}    & CNN      & 98.45    & -     & 90.21     & -     & 82.74      & -     & 85.51  & -     & 13.12    & 527.3    \\
                            & LNet$^-$    & 93.34    & -5.11 & 72.83     & -17.38 & 59.60      & -23.14 & 61.79  & -23.72 & 2.76     & 205.66   \\
                            & LNet     & 96.55    & -1.90 & 78.89     & -11.32 & 66.77      & -15.97 & 70.21  & -15.30 & 4.94     & 194.58   \\
                            & LNet-SKD & \textbf{98.66}    & +0.21  & \textbf{95.22}   & +5.01 & \textbf{85.68}      & +2.94 & \textbf{89.03}  & +3.52 & 4.94     & 194.58   \\ \midrule
\multirow{4}{*}{CICIDS2017} & CNN      & 99.87    & -     & \textbf{97.03}    & -     & 92.44      & -     & 94.32  & -     & 13.19    & 632.66   \\
                            & LNet$^-$    & 97.93    & -1.94 & 82.52     & -14.51 & 62.60      & -29.84 & 68.40  & -25.92 & 2.83     & 245.78   \\
                            & LNet     & 98.30    & -1.57 & 96.16     & -0.87 & 70.21      & -22.23 & 73.79  & -20.53 & 5.00     & 233.42   \\
                            & LNet-SKD & \textbf{99.89}    & +0.02 & 96.60     & -0.43  & \textbf{96.89}      & +4.45  & \textbf{96.74}  & +2.42  & 5.00     & 233.42   \\ \bottomrule
\end{tabular}
\label{Tab.MFM}
\end{table*}

\begin{figure*}[!t]
\centering
\subfigure[CNN]{
\includegraphics[height=3.85cm]{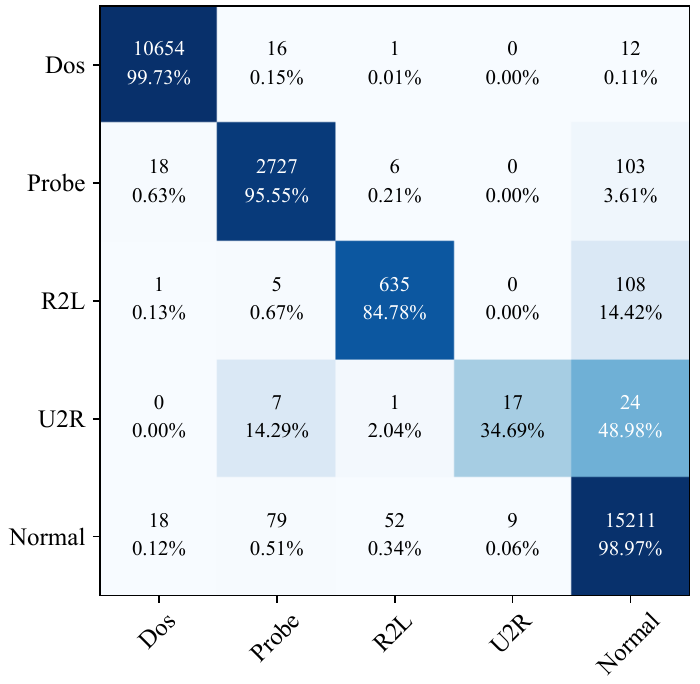} }
\hspace{0.2ex}	
\subfigure[LNet$^{-}$]{
\includegraphics[height=3.85cm]{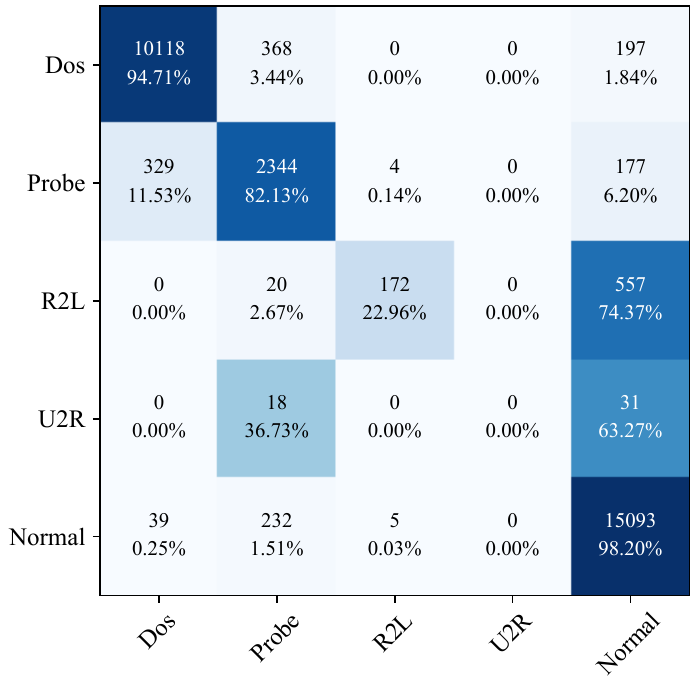} }
\hspace{0.2ex}	
\subfigure[LNet]{
\includegraphics[height=3.85cm]{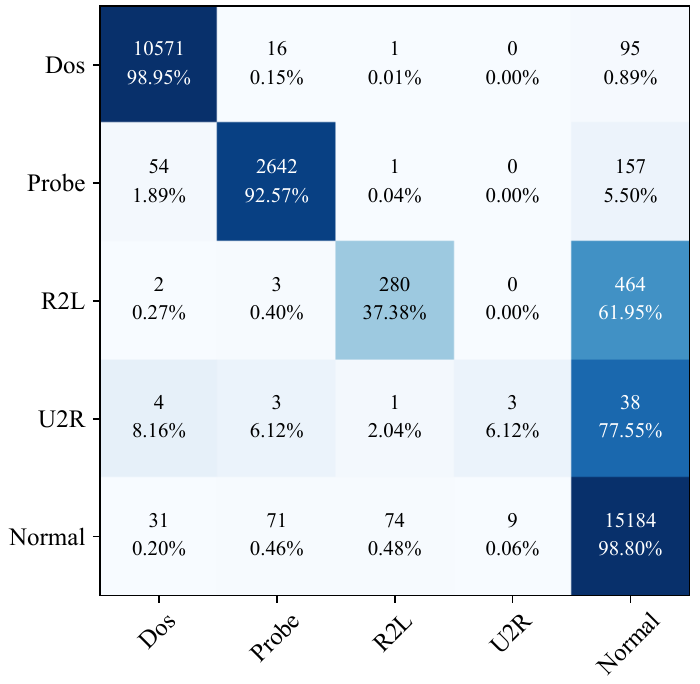} }
\hspace{0.2ex}	
\subfigure[LNet-SKD]{
\includegraphics[height=3.85cm]{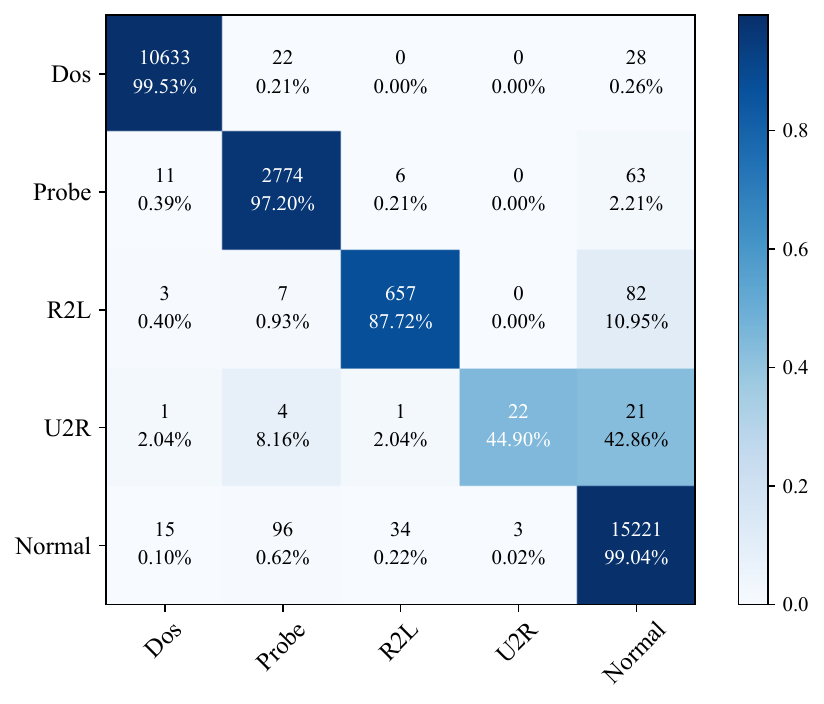} }
\caption{Confusion matrix for each model on NSL-KDD.}
\label{fig.CF}
\end{figure*}

\subsection{Evaluation Metrics}

Accuracy, Precision, Recall, and F1 score are used to evaluate the detection ability and stability of LNet-SKD. Notably, we apply macro metrics instead of micro metrics because the former is more suitable for multi-class tasks. Besides, we evaluate the computational cost via Floating Point Of Operations (FLOPs) and the number of parameters to compare the required implementation resources.

\subsection{Ablation w.r.t. Hyper-parameters}

In LNet-SKD, we take trade-offs about the performance of self-knowledge distillation with temperature parameter $\tau$ and balancing coefficient $\lambda$. Thus, effective hyper-parameters must be found to get the best classification result. Fig.~\ref{hp} shows the impact of different values for $\tau$ and  $\lambda$ used in LNet-SKD.

We fix $\lambda$ to 1 and vary $\tau$ from 1 to 10. As shown in Fig.~\ref{T}, LNet-SKD achieves the best accuracy in both datasets with a temperature $\tau = 3$, which means that more knowledge can be transferred when $\tau =3$. The effect of balance factor $\lambda$ is plotted in Fig.~\ref{L}, with a fixed $\tau = 3$. As noted in Eq. \ref{loss}, the balance coefficient $\lambda$ is considered for the contribution of the gap between two adjacent batches to the final loss. For NSL-KDD dataset, LNet-SKD performs best with $\lambda = 2$, and for CICIDS2017, the value is $\lambda = 1$.

\subsection{Ablation w.r.t. LNet-SKD}
To evaluate our proposed method comprehensively, we perform multi-classification tasks on CNN, LNet, and LNet-SKD. CNN applies a network structure similar to LNet but uses standard convolution. Moreover, a specially designed LNet$^-$ (LNet without MFM) is considered for comparison. We apply the parameters suggested by the previous experiment to train the LNet-SKD model. The result is shown in Table~\ref{Tab.MFM}.

Compared with the standard CNN model, LNet uses 37.7\% of its parameters, resulting in a succinct model with the FLOPs of 194.58K ($\downarrow$ 63.1\%). Meanwhile, our self-distillation approach does not introduce additional parameters while improving the accuracy of LNet by 2.1\% / 1.6\% on NSL-KDD and CICIDS2017, respectively, so that LNet-SKD achieves a classification accuracy on par with CNN and outperforms on F1-score. From the Table \ref{Tab.MFM}, we observe that the DeepMax block only reduces accuracy by 1.89\% and 1.57\% on two datasets compared with standard convolution. However, the value of LNet$^-$ is 5.11\% / 1.94\%, which means LNet has benefited a lot from the more representative features extracted by the DeepMax block that integrates MFM.

\subsection{Visualization}
The visualized confusion matrix in Fig.~\ref{fig.CF} further verifies the effectiveness of LNet-SKD in detecting different attack types. Especially to U2R, which is totally neglected by LNet$^-$, whereas LNet detects a few instances from more valuable features. Further enabled by SKD, LNet-SKD achieves a highly competitive detection rate even compared to the standard CNN model. It can be concluded that our approach achieves satisfactory intrusion detection performance with low model storage and computational cost.

\subsection{Compared with other methods}
\begin{table*}[!t]
\centering
\caption{Comparison of performance again other methods.}
\begin{tabular}{c|l|cc|cc|cc|cc|cc}
\toprule
Dataset                     & \multicolumn{1}{c|}{Model} & Acc.(\%) & $\Delta$(\%)     & Prec.(\%) & $\Delta$(\%)     & Recall(\%) & $\Delta$(\%)     & F1(\%) & $\Delta$(\%)     & Para.(K) & FLOPs(K)   \\ \midrule
\multirow{5}{*}{NSL-KDD}    & DNN                        & 94.23        & -     & 77.05         & -     & 62.19      & -      & 66.09        & -      & 7.17     & 14.08  \\
                            & RNN                        & 95.75        & +1.52 & 72.81         & -4.24 & 66.36      & +3.17  & 68.20        & +2.11  & 23.81    & 48.0   \\
                            & DBN\cite{DBN}              & 95.88        & +1.65 & 72.37         & -4.68 & 70.39      & +8.20  & 71.25        & +5.16  & 30.77    & 46.8       \\
                            & KD-TCNN\cite{KD-TCNN}      & 98.20        & +3.97 & 82.75         & +5.70 & \textbf{93.73}      & +31.54 & 86.63        & +20.54 & 4.62     & 358.5  \\
                           & LNet-SKD                    & \textbf{98.66}        & +4.43 & \textbf{95.22}         & +18.17 & 85.68   & +23.49 & \textbf{89.03}        & +22.94 & 4.94     & 194.58 \\ \midrule
\multirow{5}{*}{CICIDS2017} & DNN                        & 98.10        & -     & 82.82         & -     & 66.90      & -      & 71.74        & -      & 7.75     & 15.23  \\
                            & RNN                        & 98.11        & +0.01 & 82.82         & +0.00 & 68.24      & +1.34  & 73.17        & +1.43  & 24.39    & 24.58  \\
                            & DBN\cite{DBN}              & 99.84        & +1.74 & 88.75         & +5.93 & \textbf{99.49 }     & +32.59 & 92.75        & +21.01 & 31.8     & 48.0       \\
                            & KD-TCNN\cite{KD-TCNN}      & 99.64        & +1.54 & \textbf{97.43}         & +14.61 & 94.71     & +27.81 & 96.02        & +24.28 & 4.69     & 428.46 \\
                            & LNet-SKD                   & \textbf{99.90}       & +1.80 & 96.60        & +13.78 & 96.89      & +29.99 & \textbf{96.74}        & +25.00 & 5.00     & 233.43 \\ \bottomrule
\end{tabular}
\label{Tab.performance}
\end{table*}
This section provides a comparison of the LNet-SKD with baseline DL models and SOTA methods\cite{KD-TCNN, DBN} applied to the NSL-KDD and CICIDS2017 datasets. We present our experiment of LNet-SKD against the aforementioned ones in Table~\ref{Tab.performance}, where - means the calculation is not considered. As shown in the table, our LNet-SKD obtains the best performance in terms of accuracy and F1 score on both datasets with only $~4.94K$ and $~5K$ parameters, respectively. Compared with KD-TCNN designed for resource-constrained IoT devices, the LNet-SKD model has a better detection performance and lower FLOPs ($\downarrow$ 46\%). In addition, our LNet-SKD is consistent on different IDS datasets, while the other methods show various preferences in different scenarios. Hence, LNet-SKD not only achieves the trade-off between accuracy and efficiency but also is more robust and generalized for IDS.

\section{Conclusion}
This paper proposed a lightweight intrusion detection approach for resource-limited edge devices, namely LNet-SKD. We compare our model with a series of networks that applied the same architecture as LNet, as the results show, our proposed approach realizes a better trade-off between efficiency and accuracy. Specifically, LNet-SKD successfully reduces the amount of computational cost and parameter size by about 62\% with a slight improvement in accuracy and F1 score. Furthermore, LNet-SKD outperforms baseline models and other existing intrusion detection models, which is by far the state-of-the-art result with such a low resource requirement. It can be concluded that our proposed approach realizes significant superiority in network intrusion detection.

\bibliographystyle{IEEEtran}
\bibliography{reference}

\begin{thebibliography}{10}
\providecommand{\url}[1]{#1}
\csname url@samestyle\endcsname
\providecommand{\newblock}{\relax}
\providecommand{\bibinfo}[2]{#2}
\providecommand{\BIBentrySTDinterwordspacing}{\spaceskip=0pt\relax}
\providecommand{\BIBentryALTinterwordstretchfactor}{4}
\providecommand{\BIBentryALTinterwordspacing}{\spaceskip=\fontdimen2\font plus
\BIBentryALTinterwordstretchfactor\fontdimen3\font minus
  \fontdimen4\font\relax}
\providecommand{\BIBforeignlanguage}[2]{{%
\expandafter\ifx\csname l@#1\endcsname\relax
\typeout{** WARNING: IEEEtran.bst: No hyphenation pattern has been}%
\typeout{** loaded for the language `#1'. Using the pattern for}%
\typeout{** the default language instead.}%
\else
\language=\csname l@#1\endcsname
\fi
#2}}
\providecommand{\BIBdecl}{\relax}
\BIBdecl

\bibitem{networksecurity}
D.~Zaripova \emph{et~al.}, ``Network security issues and effective protection
  against network attacks,'' \emph{International Journal on Integrated
  Education}, vol.~4, no.~2, pp. 79--85, 2021.

\bibitem{surveyappication}
A.~Khraisat, I.~Gondal, P.~Vamplew, and J.~Kamruzzaman, ``Survey of intrusion
  detection systems: techniques, datasets and challenges,''
  \emph{Cybersecurity}, vol.~2, no.~1, pp. 1--22, 2019.

\bibitem{surveytype}
Z.~Ahmad, A.~Shahid~Khan, C.~Wai~Shiang, J.~Abdullah, and F.~Ahmad, ``Network
  intrusion detection system: A systematic study of machine learning and deep
  learning approaches,'' \emph{Transactions on Emerging Telecommunications
  Technologies}, vol.~32, no.~1, p. e4150, 2021.

\bibitem{CNN}
K.~Wu, Z.~Chen, and W.~Li, ``A novel intrusion detection model for a massive
  network using convolutional neural networks,'' \emph{Ieee Access}, vol.~6,
  pp. 50\,850--50\,859, 2018.

\bibitem{bilstm}
Y.~Imrana, Y.~Xiang, L.~Ali, and Z.~Abdul-Rauf, ``A bidirectional lstm deep
  learning approach for intrusion detection,'' \emph{Expert Systems with
  Applications}, vol. 185, p. 115524, 2021.

\bibitem{DBN}
O.~Belarbi, A.~Khan, P.~Carnelli, and T.~Spyridopoulos, ``An intrusion
  detection system based on deep belief networks,'' in \emph{International
  Conference on Science of Cyber Security}.\hskip 1em plus 0.5em minus
  0.4em\relax Springer, 2022, pp. 377--392.

\bibitem{zhaoIOTJ}
R.~Zhao, G.~Gui, Z.~Xue, J.~Yin, T.~Ohtsuki, B.~Adebisi, and H.~Gacanin, ``A
  novel intrusion detection method based on lightweight neural network for
  internet of things,'' \emph{IEEE Internet of Things Journal}, 2021.

\bibitem{light2}
R.~V. Mendonca, J.~C. Silva, R.~L. Rosa, M.~Saadi, D.~Z. Rodriguez, and
  A.~Farouk, ``A lightweight intelligent intrusion detection system for
  industrial internet of things using deep learning algorithms,'' \emph{Expert
  Systems}, vol.~39, no.~5, p. e12917, 2022.

\bibitem{light1}
J.-S. Pan, F.~Fan, S.-C. Chu, H.-Q. Zhao, and G.-Y. Liu, ``A lightweight
  intelligent intrusion detection model for wireless sensor networks,''
  \emph{Security and Communication Networks}, vol. 2021, 2021.

\bibitem{mobilenets}
A.~G. Howard, M.~Zhu, B.~Chen, D.~Kalenichenko, W.~Wang, T.~Weyand,
  M.~Andreetto, and H.~Adam, ``Mobilenets: Efficient convolutional neural
  networks for mobile vision applications,'' \emph{arXiv preprint
  arXiv:1704.04861}, 2017.

\bibitem{MFM}
X.~Wu, R.~He, Z.~Sun, and T.~Tan, ``A light cnn for deep face representation
  with noisy labels,'' \emph{IEEE Transactions on Information Forensics and
  Security}, vol.~13, no.~11, pp. 2884--2896, 2018.

\bibitem{KD}
G.~Hinton, O.~Vinyals, J.~Dean \emph{et~al.}, ``Distilling the knowledge in a
  neural network,'' \emph{arXiv preprint arXiv:1503.02531}, vol.~2, no.~7,
  2015.

\bibitem{zhaoKD}
R.~Zhao, Y.~Chen, Y.~Wang, Y.~Shi, and Z.~Xue, ``An efficient and lightweight
  approach for intrusion detection based on knowledge distillation,'' in
  \emph{ICC 2021-IEEE International Conference on Communications}.\hskip 1em
  plus 0.5em minus 0.4em\relax IEEE, 2021, pp. 1--6.

\bibitem{KD-TCNN}
Z.~Wang, Z.~Li, D.~He, and S.~Chan, ``A lightweight approach for network
  intrusion detection in industrial cyber-physical systems based on knowledge
  distillation and deep metric learning,'' \emph{Expert Systems with
  Applications}, p. 117671, 2022.

\bibitem{skd-batch}
Y.~Shen, L.~Xu, Y.~Yang, Y.~Li, and Y.~Guo, ``Self-distillation from the last
  mini-batch for consistency regularization,'' in \emph{Proceedings of the
  IEEE/CVF Conference on Computer Vision and Pattern Recognition}, 2022, pp.
  11\,943--11\,952.

\bibitem{signature}
A.~Drewek-Ossowicka, M.~Pietro{\l}aj, and J.~Rumi{\'n}ski, ``A survey of neural
  networks usage for intrusion detection systems,'' \emph{Journal of Ambient
  Intelligence and Humanized Computing}, vol.~12, no.~1, pp. 497--514, 2021.

\bibitem{KNN}
H.~Ding, L.~Chen, L.~Dong, Z.~Fu, and X.~Cui, ``Imbalanced data classification:
  A knn and generative adversarial networks-based hybrid approach for intrusion
  detection,'' \emph{Future Generation Computer Systems}, vol. 131, pp.
  240--254, 2022.

\bibitem{SVMBayes}
J.~Gu and S.~Lu, ``An effective intrusion detection approach using svm with
  na{\"\i}ve bayes feature embedding,'' \emph{Computers \& Security}, vol. 103,
  p. 102158, 2021.

\bibitem{LightGBM}
J.~Liu, Y.~Gao, and F.~Hu, ``A fast network intrusion detection system using
  adaptive synthetic oversampling and lightgbm,'' \emph{Computers \& Security},
  vol. 106, p. 102289, 2021.

\bibitem{shallow}
R.~Vinayakumar, K.~Soman, and P.~Poornachandran, ``Evaluating effectiveness of
  shallow and deep networks to intrusion detection system,'' in \emph{2017
  International Conference on Advances in Computing, Communications and
  Informatics (ICACCI)}.\hskip 1em plus 0.5em minus 0.4em\relax IEEE, 2017, pp.
  1282--1289.

\bibitem{AutoRNN}
G.~D’Angelo and F.~Palmieri, ``Network traffic classification using deep
  convolutional recurrent autoencoder neural networks for spatial--temporal
  features extraction,'' \emph{Journal of Network and Computer Applications},
  vol. 173, p. 102890, 2021.

\bibitem{NSL-KDD}
M.~Tavallaee, E.~Bagheri, W.~Lu, and A.~A. Ghorbani, ``A detailed analysis of
  the kdd cup 99 data set,'' in \emph{2009 IEEE symposium on computational
  intelligence for security and defense applications}.\hskip 1em plus 0.5em
  minus 0.4em\relax Ieee, 2009, pp. 1--6.

\bibitem{CICIDS}
I.~Sharafaldin, A.~H. Lashkari, and A.~A. Ghorbani, ``Toward generating a new
  intrusion detection dataset and intrusion traffic characterization.''
  \emph{ICISSp}, vol.~1, pp. 108--116, 2018.

\end{thebibliography}
\end{document}